\documentclass[11pt,a4paper]{article}
\usepackage{geometry}
\usepackage[round]{natbib}
\usepackage{authblk}
\usepackage{graphicx}
\usepackage{tabularx}
\usepackage{multirow}
\usepackage{multicol}
\usepackage{amsthm}
\usepackage{algorithm2e}
\usepackage{ulem}
\usepackage{color}
\usepackage{amsmath}
\usepackage{cite}
\usepackage[caption=false]{subfig}

 \usepackage{lipsum}

\usepackage{ragged2e}
\usepackage{sectsty}
\sectionfont{\fontsize{15}{15}\selectfont} 
\subsectionfont{\fontsize{13}{15}\selectfont}
\usepackage{tikz} 
\usetikzlibrary{shapes,arrows}

\usepackage{setspace}

 \newgeometry{a4paper, total={170mm,257mm},left=25mm,right=25mm,bottom=25mm,top=25mm,}

\begin{document}
\pagenumbering{gobble}
\title{\LARGE \textbf{Designing wildlife crossing structures for ungulates in a desert landscape: A case study in China}}
\centering
\author [1] { Bin Zhang}
\author[2,3]{ Junqing Tang\thanks{Co-first author}}
\author [3]{Yi Wang}
\author[4]{Hongfeng Zhang}
\author [5] {Gang Xu}
\author[5]{Yu Lin}
\author[4]{Xiaomin Wu\thanks{Corresponding author at: xiaominwu66@126.com}}

\affil [1]  { \textit{Gansu Yuanda Group Ltd., Lanzhou, Gansu, China}}
\affil[2]{ \textit{School of Civil Engineering \& Mechanics, Huazhong University of Science \& Technology, Wuhan, 430074, China}}
\affil[3]{ \textit{ETH Zurich, Future Resilient Systems, Singapore-ETH Centre, 1 CREATE Way, CREATE Tower, 138602 Singapore}}
\affil [4] { \textit{ Shaanxi Institute of Zoology, Xi'an, Shaanxi, China}}
\affil[5]{ \textit{Tianjin Research Institute for Water Transport Engineering, Ministry of Transport of China, Tianjin, China}}
\renewcommand\Authands{ }
\renewcommand\Authsep{  }
\renewcommand\Authfont{\bfseries}

\date{}
\maketitle

\begin{abstract}

This paper reports on the design of wildlife crossing structures (WCSs) along a new expressway in China, which exemplifies the country's increasing efforts on wildlife protection in infrastructure projects. The expert knowledge and field surveys were used to determine the target species in the study area and the quantity, locations, size, and type of the WCSs. The results on relative abundance index and encounter rate showed that the ibex (\textit{Capra ibex}), argali sheep (\textit{Ovis ammon}), and goitered gazelle (\textit{Gazella subgutturosa}) are the main ungulates in the study area. Among them, the goitered gazelle is the most widely distributed species. WCSs were proposed based on the estimated crossing hotspots. The mean deviation distance between those hotspots and their nearest proposed WCSs is around 341m. In addition, those 16 proposed underpass WCSs have a width of no less than 12m and height of no lower than 3.5m, which is believed to be sufficient for ungulates in the area. Given the limited availability of high-resolution movement data and wildlife-vehicle collision data during road's early design stage, the approach demonstrated in this paper facilitates practical spatial planning and provides insights into designing WCSs in a desert landscape.

\vspace{0.5cm}

\raggedright
\text{\textit{Keywords:}}
Road ecology; Wildlife crossing structures; Desert landscapes; Ungulates; Underpasses.
\end{abstract}

\setlength\parindent{2em}\justify

 \newgeometry{a4paper, total={170mm,257mm},left=30mm,right=30mm,bottom=30mm,top=30mm,}
\pagenumbering{arabic}
\setlength\parindent{2em}\justify 


\section{Introduction}
Conflicts between expanding infrastructure and preserving regional biodiversity have been increasingly prominent in countries such as China in recent years~\citep{wang2015china}. Linear transportation infrastructure, such as expressways, can cause habitat fragmentation and genetic isolation, imposing negative environmental and ecological impacts on landscapes and biodiversity~\citep{forman2003road,corlatti2009ability}. Therefore, mitigation measures, such as Wildlife Crossing Structures (WCSs), are needed to compensate these deleterious consequences~\citep{de2018landscape}.

Here, we report on a recent road project in which we practically designed WCSs along a segment of the Beijing-Urumqi G7 national expressway in Gansu Province. G7 expressway is known as one of the recent infrastructure milestones in China, whose considerable portion lies in China's northwestern desert region. Among its provincial segments, the new segment in Gansu Province has an indispensable role as a vital connector along China's new ``Silk Road" within the ``Belt and Road Initiatives"~\citep{BeltRoad}. The region also features a provincial nature reserve with diverse wildlife species. Particularly, it is known as an essential habitat for several valuable ungulate species, including the ibex (\textit{Capra ibexthe}), goitered gazelle (\textit{Gazella subgutturosa}), wild Bactrian camel (\textit{Camelus ferus}), and Mongolian kulan (\textit{Equus hemionus ssp. hemionus}). Moreover, due to human activities, the latter two have become extremely rare in the area. Therefore, it is essential to implement WCSs along this new segment of the G7 expressway.

Two common types of WCS, overpasses and underpasses, can provide linkages, improve connectivity, and mitigate the incidence of roadkills~\citep{pierik2016designing,ernst2014quantifying}. Overpasses, or ``green bridges"~\citep{glista2009review}, provide open views and sufficient space for animals, which is generally more suitable for species that instinctively avoid narrow and dim spaces. On the other hand, underpasses can be categorized as small, intermediate, large, or very large. Those within the first classification are intended for small mammals, amphibians, and reptiles, often taking the form of tunnels or small culverts. Intermediate underpasses usually serve for both crossing and water drainage~\citep{grilo2010mitigation}. The target species for this category are small and medium-sized animals such as rodents, fox, coyote, and ocelot. The last two types of underpasses are the most commonly applied types of WCS and provide effective crossings for a wide range of species~\citep{grilo2010mitigation}. More importantly, they can be effectively incorporated into standard engineering structures, such as bridges, in road projects~\citep{smith2015wildlife}.

To date, a large amount of the previous research on WCS has concentrated on post-project monitoring of those mitigation measures~\citep{clevenger2000factors,bond2008temporal,olsson2008effectiveness,wang2017monitoring}, assessing their effectiveness through factor analysis~\citep{mcdonald2004elements,clevenger2005performance,ji2017impact}. More recently, attention has expanded to include various species, especially small species~\citep{goosem2000effects}, and has called for the understandings of the effects of larger spatial-temporal scales, such as species communities and ecosystems~\citep{van2011effects,van2015ecological}. In practice, although several guidelines are available for practically designing WCSs~\citep{meese2009wildlife, kautz2010wildlife, clevenger2011wildlife}, only a few case studies of WCS design prior to road construction exist in the literature. They include the WCSs along the Trans-Canada Highway in Banff National Park in Canada~\citep{Trans-Canada} and the US 93 within the Flathead Indian Reservation in Montana, USA~\citep{US93}. The literature on the practical issues of mitigation measures is also scarce regarding road planning and management in the field of road ecology research~\citep{van2015ecological,roedenbeck2007rauischholzhausen}.  

Scholars have recently raised concerns about biodiversity loss and environmental degradation during the massive expansion of infrastructure in China~\citep{ascensao2018environmental}. Even though the country began to initiate WCS practice relatively late compared with Australia, Europe, and North America, the government has acknowledged the necessity for integrating road ecology principles in infrastructure projects, which has been emphasized by the Chinese Ministry of Transport and in new national laws, e.g., the Wildlife Protection Law of the People's Republic of China~\citep{wang2015china,kong2013road}. In particular, the implementation of WCSs have been brought to prominence in the planning of the Qinghai-Tibet Railway~\citep{xia2007effect}. Years of continuous monitoring has proven the effectiveness of these structures on preserving local species, including the Tibetan antelope (\textit{Pantholops hodgsonii}) and Tibetan gazelle (\textit{Procapra picticaudata})~\citep{yang2008tibetan}. Quite a few WCSs have also been installed in other infrastructure projects in the country, such as the underpasses intended for Asian elephant (\textit{Elephas maximus}) along the Simao-Xiaomengyang highway in Yunnan Province~\citep{pan2009corridor}, as well as the overpasses built for the giant panda (\textit{Ailuropoda melanoleuca}) within the Qinling giant panda corridor~\citep{Panda}.

In general, the following two gaps are evident in those earlier studies: (1) Previous studies that focus on practically designing WCS in the early design stage of a road project are relatively less reported. Also, (2) the contributions of infrastructure projects considering wildlife preservation from Asian countries have been relatively less reported~\citep{taylor2010roads}, especially in developing countries, such as China.

To address these gaps, we conducted our study, aiming to provide a practical showcase and insights into WCS design in a desert region in northwestern China. We evaluated the target species, as well as the quantity and locations of WCSs through expert knowledge and a large amount of field surveys. The species abundance was investigated using line transects and the possible crossing hotspots were estimated through a belt transect survey. Surveys of surface waters and sign-tracking investigations were conducted to support our decision in the end. Moreover, this study exemplifies China's increasing proactive attitude in promoting environmental protection in infrastructure development in recent years.

The main contribution of this paper is summarized as follows:
\begin{enumerate}

\item This paper demonstrates China's recent effort to designate proactive wildlife protection as a high priority in an expressway project. This is of great significance because few previous publications have dealt with such design of WCSs in a desert landscape in China. 

\item The field investigations provide an up-to-date understanding of the large ungulate species in the study area, including confirming the absence of the wild Bactrian camel and Mongolian kulan.

\item The paper practically demonstrates an approach for designing WCSs when high-resolution movement data and wildlife-vehicle collision data are out of availability during road's early design stage. This approach could be recommended to other projects with similar constraints.

\end{enumerate}

The remainder of the paper is organized as follows: In Section 2, we provide essential materials and background information about the expressway project and study area. In Section 3, main methodologies are introduced, including collection of expert knowledge, techniques in field surveys, and data analysis methods. Section 4 presents the survey results to answer questions as ``For whom?" (the target species and estimated crossing hotspots). After, Section 5 presents the practical design of WCSs in this project, including ``How many?", ``Where at?" (the total number of proposed WCSs and their locations), and ``What type and size?" (proposed type and size of the WCSs). Finally, we discuss the results in Section 6 and conclude the study in Section 7.

\section{Materials}
\subsection{Project timeline}
Infrastructure projects are accomplished by following consecutive stages within the project development cycle, from inception through planning, design, construction, and finally, to operation and maintenance~\citep{roberts2015incorporating}. Our work took place during the early design stage. Simply put, the expressway was not built when we conducted our field investigations. We only have initial design details from the engineering team, such as the planned location of the expressway, road structures, and bridge designs. The main construction process of the project started in mid-2013 and completed in 2015. The entire G7 expressway, from Beijing to Urumqi, was not in use until 2017. Thus, the most difficult problem for us was the lack of relevant data, such as monitoring data of animal movement and wildlife-vehicle collision data.

\subsection{Project layout and expressway design}
\label{expressway}
\begin{figure}[h!]
\centering
\includegraphics[width=1\textwidth]{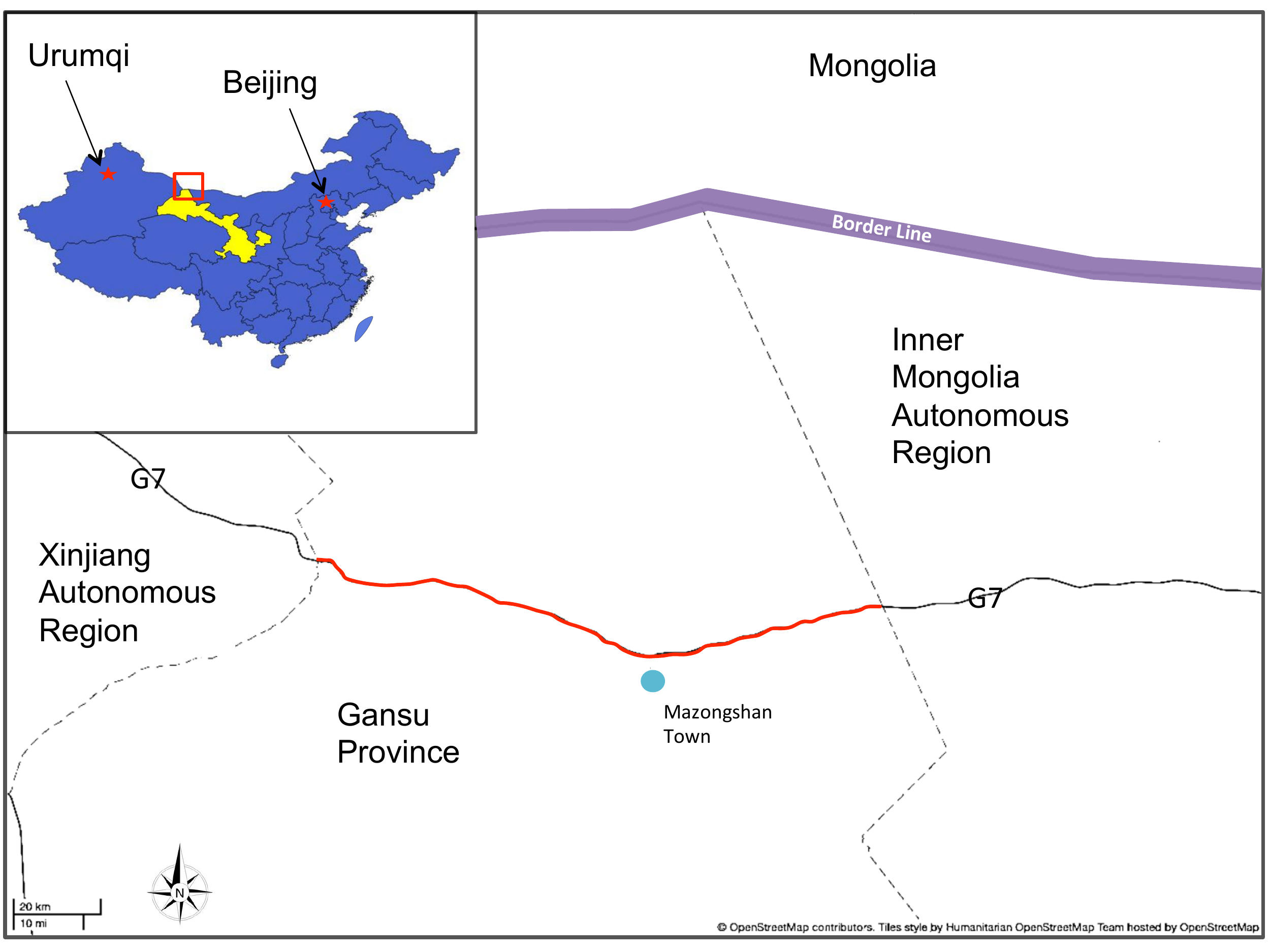}
\caption{\small Geographical layout (Source: OpenStreetMap\textcopyright). Red line, G7 alignment in the study area; bold purple line, the national border between Mongolia and China; dotted lines, the provincial boundary between provinces; In the subplot, yellow shading, Gansu Province; red box, the study area; stars, origin and destination of G7 national expressway. (color is needed)}
\label{fig_1}       
\end{figure}

The study area (Fig.~\ref{fig_1}) is located in the Mazongshan region of the Gansu Province, China. This segment of G7 is designed to start from the boundary between Gansu Province and the Inner Mongolia Autonomous Region and end at the provincial boundary of the Xinjiang Autonomous Region (136.67km in total). The expressway is designed to be a sealed, double-track, four-lane road, with a maximum design speed of 120km/h. In addition, 57 bridges (40 small, 10 medium, and 7 large) and 226 culverts were planned in the original engineering designs (see Appendix A). 

\subsection{Local- and landscape-level background}
The overall landscape of the study area primarily consists of subtropical desert and semi-arid land, with fragmented areas of grasslands and steppes. Surface rivers are rare, and the annual precipitation is only approximately 76mm, with some extremely desertified areas receiving less than 10mm each year. The underground water has a high mineralization level (i.e., 5-30g/L)~\citep{WangThe2000}. No permanent or seasonal rivers exist, and only temporary ponds and small salt marshes sparsely distribute in the area. Vegetation coverage is also very low. The main plant species are those commonly found in deserts and semi-arid landscapes, including saxaul (\textit{Haloxylon ammodendron}), ephedra (\textit{Ephedra przewalskii}), and zygophyllum (\textit{Zygophyllum xanthoxylum}). The area contains no rare or endangered plant species.

One provincial nature reserve lies approximately 40km southeast of the Mazongshan town. It is established for ibex (\textit{Capra ibex}), goitered gazelle (\textit{Gazella subgutturosa}), and argali sheep (\textit{Ovis ammon}). In total, 10 families of 35 mammalian species, 12 families of 30 bird species and six families of 12 reptile species have been documented in this region~\citep{ChenThe1997}. Table~\ref{tab3} summarizes the main species of large- and medium-sized mammals in the area, their protection class according to the new Wildlife Protection Law of People's Republic of China, and their Red List index from International Union for Conservation of Nature (IUCN). Additionally, historical records show that the wild Bactrian camel (\textit{Camelus ferus}) and the Mongolian kulan (\textit{Equus hemionus ssp. hemionus}) used to inhabit in this area but have been rarely detected since the 1980s. Nevertheless, we still considered those two species in the WCS designs.

\begin{table}[]

\caption{\small Main large- and medium-sized mammals in the study area}
\label{tab3}
\begin{tabular}{|l|l|l|l|}
\hline
\centering
 \textbf{Scientific name} & \textbf{Common name} & \textbf{Protection class} & \textbf{IUCN's} \\
 &&\textbf{in China}&\textbf{Red List} \\ \hline
  Capra ibex & ibex & I & LC \\ \hline 
  Equus hemionus ssp. hemionus & Mongolian kulan & I & NT \\ \hline 
  Camelus ferus & wild Bactrian camel & I & CR \\ \hline 
  Ovis ammon & argali sheep & II & NT \\ \hline
  Gazella subgutturosa & goitered gazelle & II & VU \\ \hline
  Lynx lynx & Eurasian lynx & II & LC \\ \hline 
  Otocolobus manul & Pallas's cat & II & NT \\ \hline
  Felis bieti & Chinese mountain cat & II & VU \\ \hline 
  Canis lupus & wolf & - & LC \\ \hline
  Vulpes corsac & corsac fox & - & LC \\ \hline
\end{tabular}

* \small{Ex-Extinct; EW-Extinct in the wild; CR-Critically endangered; EN-Endangered; VU-Vulnerable; NT-Near threatened; LC-Least concern; DD-Data deficient; NE-Not evaluated}
\end{table}

\section{Methodology}

\subsection{Collection of expert knowledge}
Professional assessments and expert knowledge are essential when there is no ready-to-use data related to animal movements or roadkills~\citep{clevenger2002gis}. Involving shareholders in a systematic decision-making process is also a common practice~\citep{baxter2018turning}. We visited several local authorities, including the Gansu Wildlife Conservation Bureau, the Provincial Administration of Nature Reserve, and several local forestry departments, to invite local shareholders to share their experiences and engage in the WCS design. We inquired about local flora and fauna (both current and historical species), hydrography, habitat landscapes, and local wildlife census. We also invited eight experts (three with backgrounds and expertise in forestry, plus one in ecology and four from highway engineering institutes/companies, each of whom had more than five years of professional experience) to participate in our field trips. Afterward, we organized symposiums to discuss the WCS designs and share opinions among the experts (10 people), delegates of local authorities and villages (12 people), and our research team (6 people).

\subsection{Field surveys}
The \textit{line transect method}, or line-intercept sampling, is a widely applied biological and ecological sampling technique for estimating the abundance of animal populations~\citep{kaiser1983unbiased}. Briefly, signs of a species will be recorded if a selected line intersects with them. Note that these signs could include tracks, feces, hair, carcasses or body parts, etc. By recording the appearance frequency of the signs, the species abundance can be determined accordingly. Another popular technique, the \textit{belt transect method}, is similar to the line transect method but investigates a study area with a belt with designed length and width~\citep{southwood2009ecological}. Because this method investigates an area, it can depict not only the abundance but also provide hints on the presence-absence information of species within the belt area. Conducting this method is also an opportunity to survey animal tracks on the roadside if belts are implemented alongside or overlapping the road, which is particularly useful for inspecting crossing hotspots along a long road~\citep{smith2015field}. 

Moreover, we conducted \textit{surface water surveys} and evaluated their overall hydrographical conditions. Most importantly, we would like to know their basic characteristics, such as hydroperiod, a term to indicate the number of months per year a pond contains water~\citep{razgour2010pond}, and investigate their impact on wildlife in the study area. There, we also applied the \textit{sign-tracking technique} to trace the direction of wildlife's movements and landscape-scale distribution patterns near the water points. The purpose of water surveys is to support the transect surveys and provide supplementary estimation on wildlife activeness in the area.

We placed line transects in an evenly distributed manner, from east to west, which were perpendicular to the planned alignment of the expressway. This resulted in 33 lines (total length of 373km) and one belt (15m wide and 137km long) to cover the full length of the expressway. For each survey trip on the lines and belt, we recorded the number of animal signs. The surveys were carried out in two sessions, three months apart. From 10 December to 24 December in 2012, we conducted a belt transect survey and recorded all the animal signs. The line transect sampling was performed from 08 March to 23 March in 2013. Overall, we investigated approximately 4000 km by a team of 16 investigators in 35 days, i.e., 30 days for transect surveys (both line and belt) plus 5 days for surveys of surface waters and landscape features.

\subsection{Data analysis}
Two ecological measurement indexes were applied for data analysis. The first is the relative abundance index (RAI), which indicates how abundant one species is with respect to all the encountered species. The RAI of species $i$ can be expressed as:

\begin{equation}
RAI_{i} (\%)= \frac{N_{i}}{T}\times 100
\label{eq1}
\end{equation}
where $N_{i}$ is the number of signs counted for the species $i$ and $T$ is the total number of signs for all the encountered species.

The second index is the encounter rate of one species (ER) during all the investigations along one line transect. This indicator represents how likely the signs of one species can be encountered per unit distance, and is calculated as:

\begin{equation}
ER_{i,j} (\%)= \frac{N_{i}}{L_{j}}\times 100
\label{eq2}
\end{equation}
where $ER_{ij}$ is the encounter rate of species $i$ along line $j$, $L_{j}$ is the total distance covered in line $j$. These two indices provided an understanding about the current state of wildlife species in the study area, as well as an estimation of the primary target species for which the WCSs should be designed.

\section{Field survey results}
\subsection{Target species}
The RAI and ER values for each species can be calculated using the data obtained from line transect surveys. The result of RAI can identify the target species from an overall perspective, i.e., what the relatively abundant ones are. Besides, the result of ER can further reveal species-specific abundance on each line and overall animal distribution in the study area. Note that, during the survey, signs from multiple species were found. Apart from large- and medium-sized mammals, there are also small mammals, such as the tolai hare (or Mongolian brown hare;  \textit{Lepus tolai}) and corsac fox (\textit{Vulpes corsac}). However, only the signs of large- and medium-sized species were recorded because of their high requirements on crossing structures. 

\begin{figure}[h!]
\centering
\includegraphics[width=0.8\textwidth]{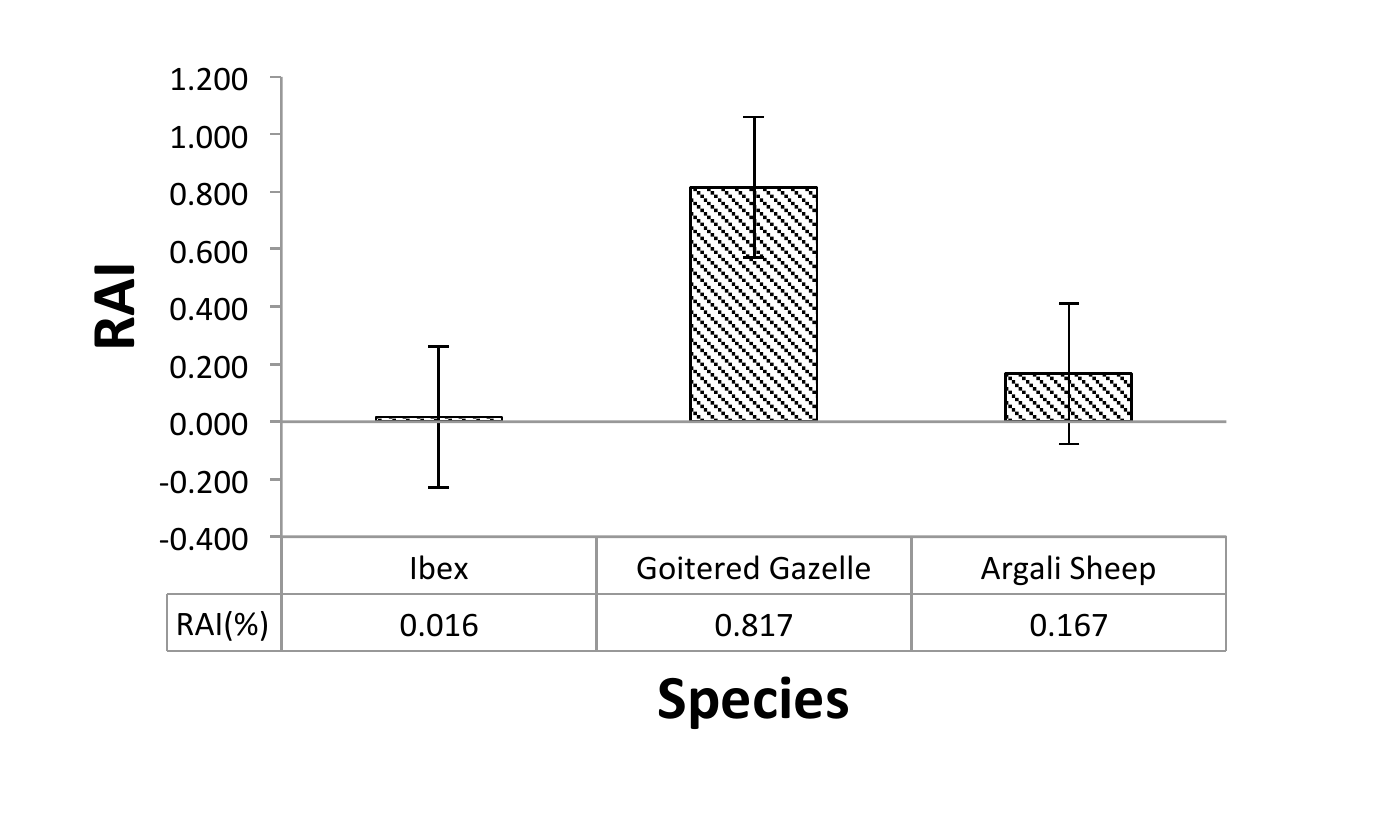}
\caption{\small Relative Abundance Index (RAI) of observed species based on survey data. (color is NOT needed)}
\label{Fig5}       
\end{figure}

Fig.~\ref{Fig5} shows the result of overall RAI obtained from surveys for three main ungulate species in the area, including ibex (1.6\%), goitered gazelle (81.7\%) and argali sheep (16.7\%). We found a large number of signs of goitered gazelle during surveys. This result is in line with the historical wildlife census, showing that the goitered gazelle has been the most abundant ungulate species in the study area.

\begin{figure}[h!]
\centering
\includegraphics[width=0.8\textwidth]{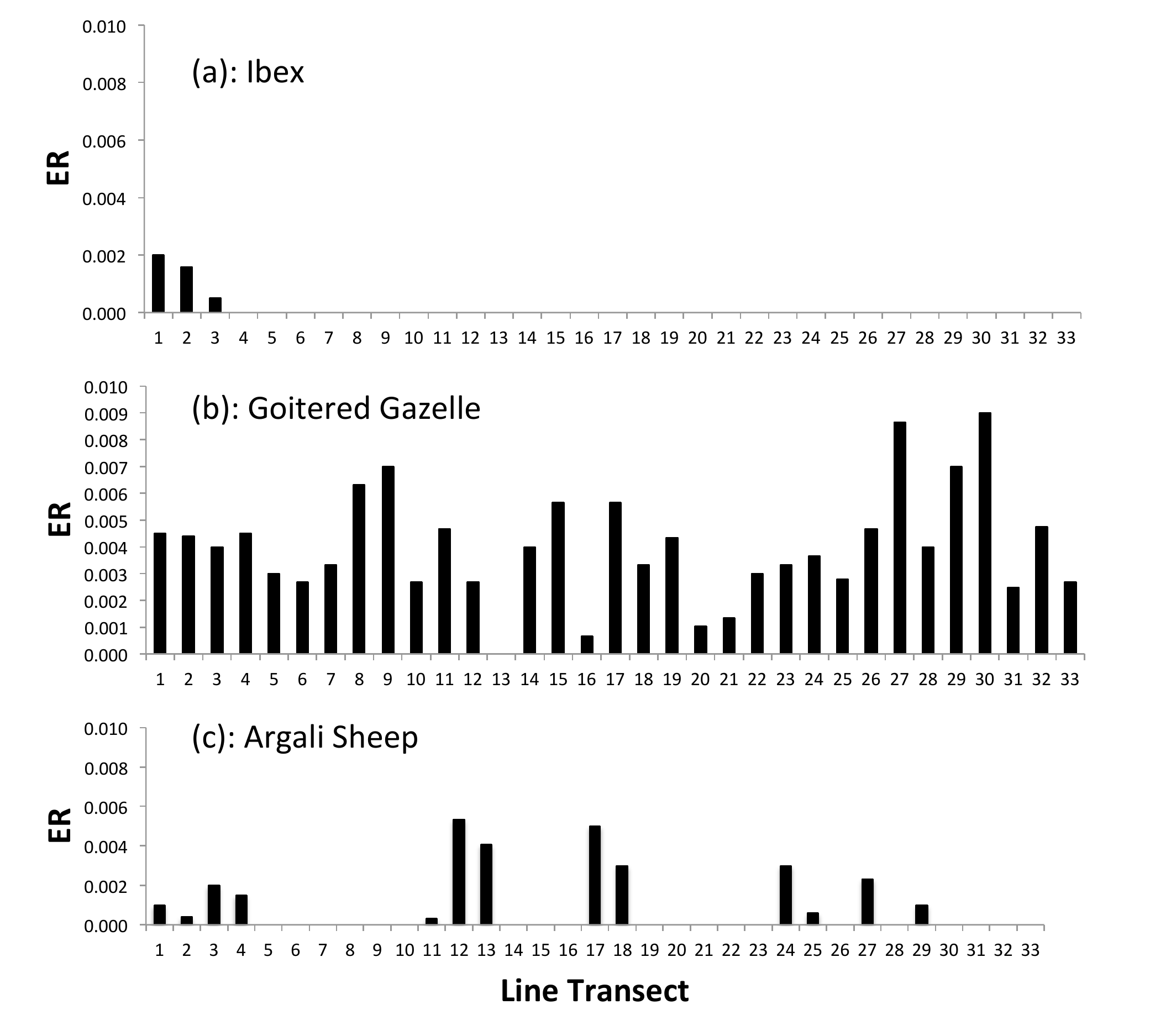}
\caption{\small Encounter Rate (ER) of observed species based on survey data. (color is NOT needed)}
\label{Fig4}       
\end{figure}

Fig.~\ref{Fig4} presents the result of ER on each line transect for these three ungulate species. As can be seen from Fig.~\ref{Fig4} (a), the ibex was only detected along the first three line transects during surveys, and the ER values were all less than 0.0025 encounters/km which is relatively low. On the rest of the line transects, we found no more signs of this species. The ER values for the argali sheep was also relatively low as shown in Fig.~\ref{Fig4} (c). A significant number of signs of this species were detected only on transects 12 to 13 and 17 to 18 (around 0.003 to 0.006 encounters/km). The rest of the signs were mostly spotted on transects 3, 4, 24, and 27, with the ER values of less than 0.004 encounters/km. In contrast, as can be seen in Fig.~\ref{Fig4} (b), goitered gazelles have greater abundance and wider distribution than the other two species do in the area. Signs of this species were observed on almost every line transect and the ER values reach 0.006 to 0.010 encounters/km on the 8, 9, 27, 29 and 30 line transects.

\subsection{Estimated crossing hotspots}
\label{belt survey}
\begin{figure}[h!]
\centering
\includegraphics[width=1\textwidth]{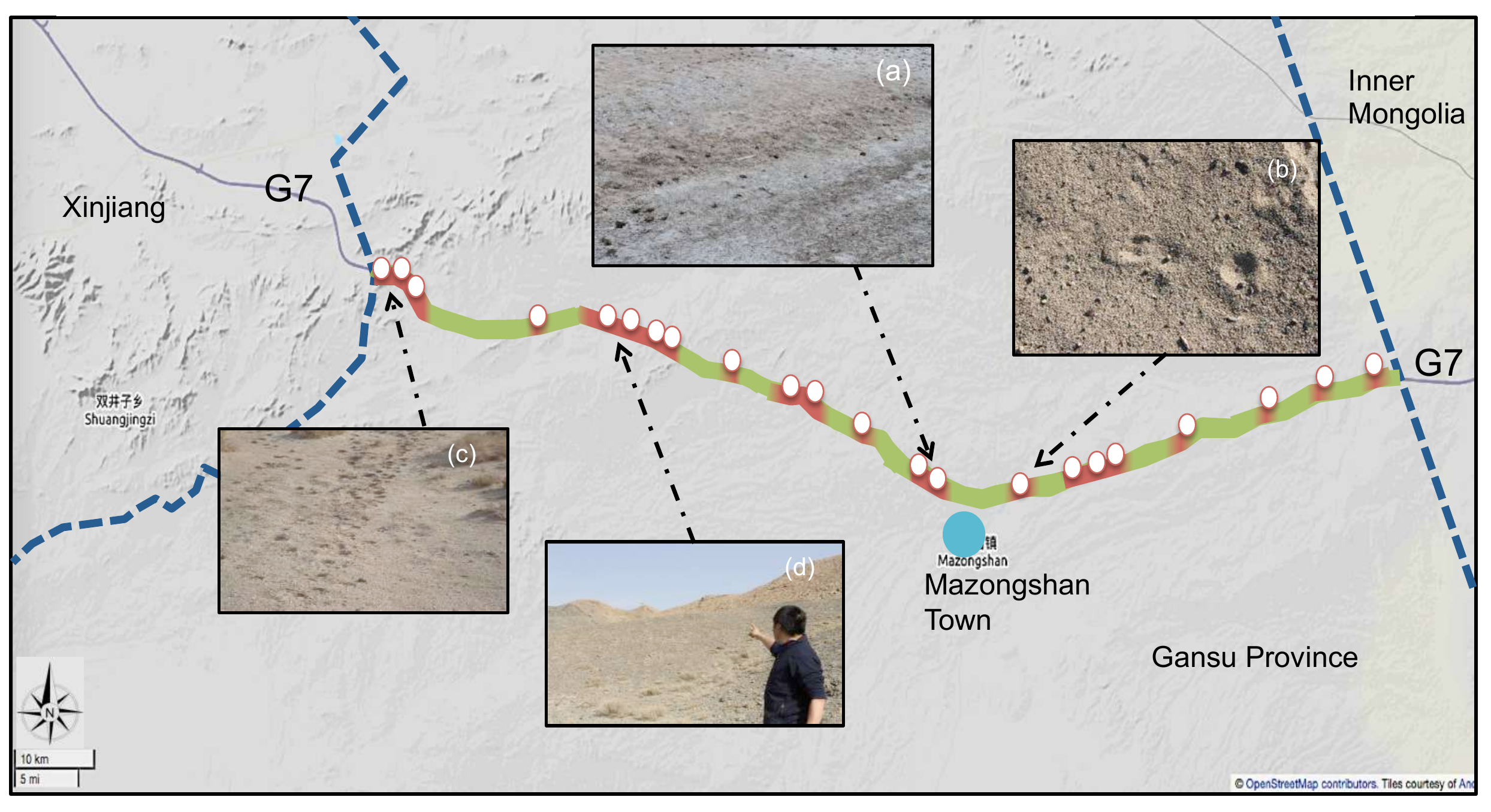}
\caption{\small Estimated crossing hotspots according to animal signs found in belt transect survey. Photos are examples of some selected field observations: (a) Footprints and excrement of goitered gazelle; (b) Footprints and excrement of goitered gazelle; (c) Footprints of goitered gazelle; (d) One argali sheep. Note that the western areas are mountainous terrains (Adapted from OpenStreetMap~\textcopyright). (color is needed)}
\label{Fig6}       
\end{figure}

The results from the belt transect survey showed the existence of animal signs near the planned location of the expressway, which suggested active wildlife movements around it. We estimated 22 crossing hotspots based on the signs (as shown in Fig.~\ref{Fig6}), which are non-uniformly located along the expressway. We found that, in the east, the hotspots were relatively sparsely located with a mean interval of 8km. In the middle portion, this value reduced to approximately 5km. As approaching west, there was a distinct section at where four hotspots were relatively crowded (with intervals less than 2km). After that, a much larger interval of more than 10km was found before the last three hotspots, while the distance intervals among the last three hotspots were much shorter (only around 2km on average). Moreover, the belt transect survey confirmed our findings on the target species from the line transect surveys, that is, most of the signs were from goitered gazelles. Unfortunately, the presence-absence information from the survey additionally confirmed the absence of the wild Bactrian camel and Mongolian kulan. 


\subsection{Water surveys}
Surface waters are vital attracting points for animals to fulfill their daily water needs and, therefore, affect the distribution patterns of wildlife in an arid environment~\citep{roshier2008animal,razgour2010pond,zhang2015water,kluever2017influence}. As indicated by the survey result, there were no permanent rivers in this area and only seven small temporary ponds were found. Fig~\ref{Fig7} in Appendix B summarizes all seven ponds and their perpendicular distance to the planned location of the expressway. We found abundant animal footprints at each pond. While most of the footprints were from ungulates, we also observed signs of small rodents, wolves, and even birds. It is evident that, as primary water supply points in the area, these surface waters have high attractive strength for a wide range of wildlife. By tracking the footprint trajectories using sign-tracking techniques, we roughly estimated the direction of animal movements and found that the ungulate species have a vast home range in the area (the range used for their daily activities).

\section{Design of crossing structures}
\label{line graph}
\begin{figure}[h!]
\centering
\includegraphics[width=1\textwidth]{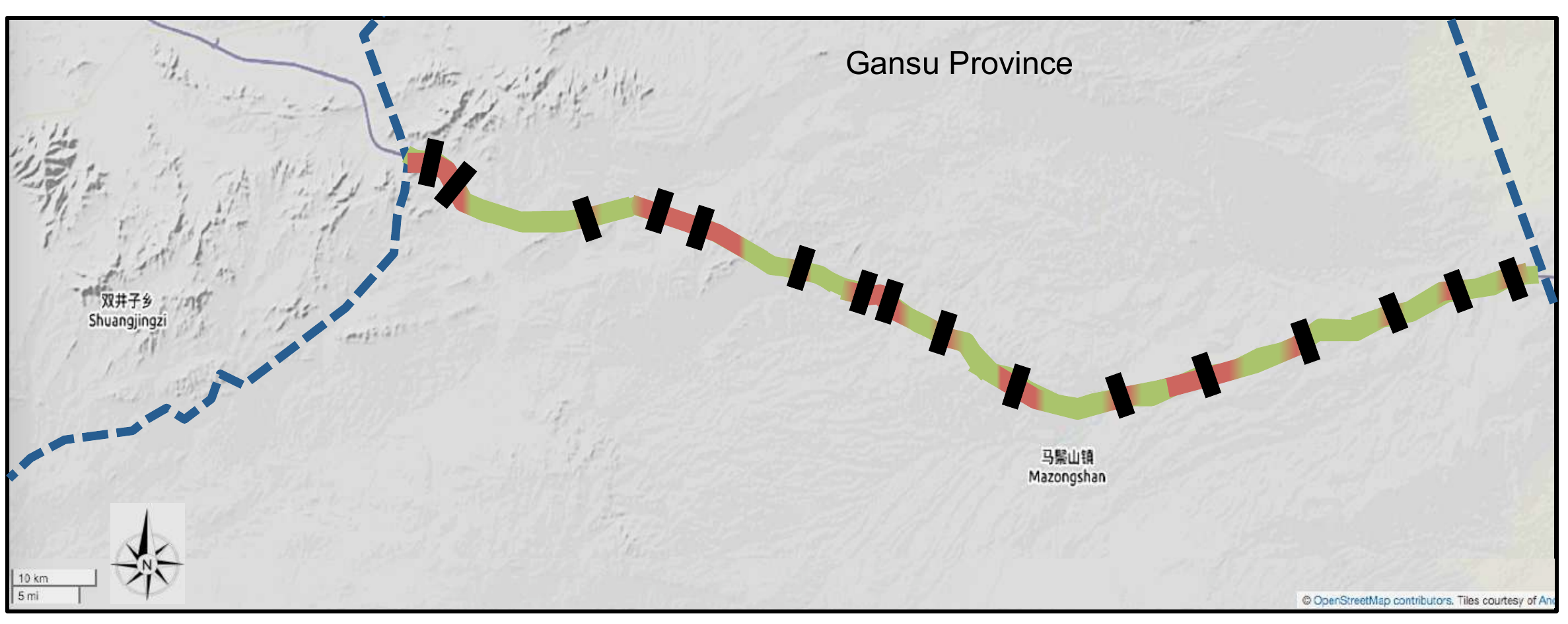}
\caption{\small Proposed locations for 16 WCSs along planned expressway and the hotspots identified from the belt transect survey (Map is adapted from OpenStreetMap~\textcopyright). (color is needed)}
\label{Fig8}       
\end{figure}

The underpass was proposed as the type of WCSs for this project. Based on survey results and expert knowledge, we proposed 16 locations along the expressway that could be suitable for implementing WCSs, as shown in Fig.~\ref{Fig8}. We double-checked the engineering design and found that the planned bridge structures could cover 15 out of those 16 proposed WCSs so that only one new needs to be added.

\begin{figure}[h!]
\centering
\includegraphics[width=1\textwidth]{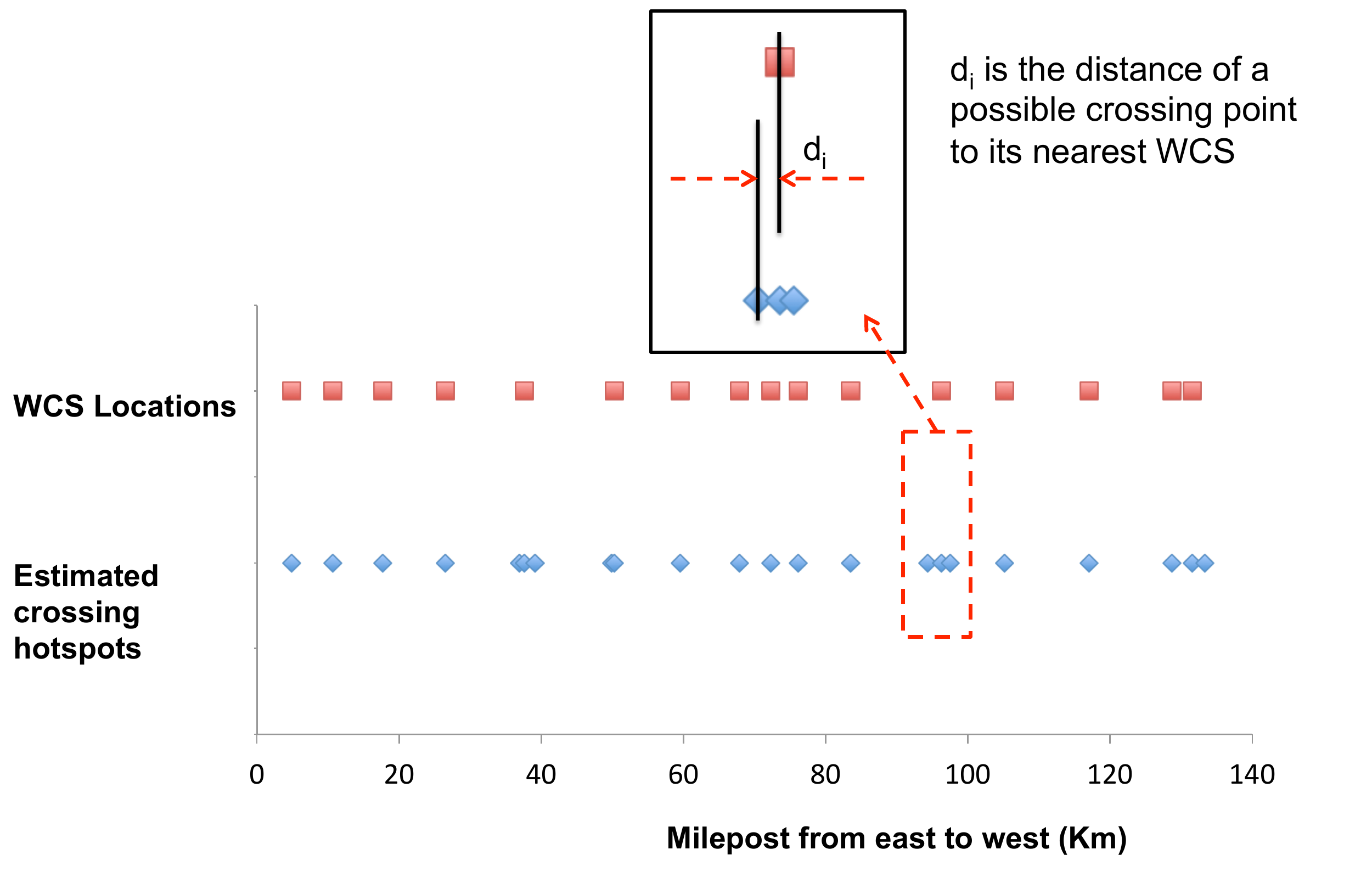}
\caption{\small Comparisons between WCS locations and estimated crossing hotspots. Highlighted subplot illustrates the definition of deviation distance between one estimated hotspot and its closest WCS. (color is NOT needed)}
\label{Fig9}       
\end{figure}

Fig.~\ref{Fig9} shows the location comparison between those 16 proposed WCSs and the 22 hotspots estimated from the belt survey. As can be seen that most of the proposed WCSs are located at the very vicinity of a hotspot, except around several crowded hotspots, only one WCS is proposed. We then calculated the deviation distance, $d_{i}$ as illustrated in the subplot, between the estimated crossing hotspots and their closest WCSs. The mean value of $d_{i}$ is 341m only. 

When deciding on the appropriate size of these underpasses, the consideration for the timidity and cautionary habits of ungulate species is of vital importance. Therefore, sufficient space and brightness are pivotal. Based on manuals~\citep{meese2009wildlife, kautz2010wildlife} as well as expert knowledge, we proposed that the dimension of these WCSs can be considered as: (1) The overall width of each structure should be no less than 12m, and (2) The height should be no less than 3.5m, wherein the heights for specific species being $\geq$ 4.5m for the Mongolian kulan and wild Bactrian camel, $\geq$ 4.0m for the ibex and argali sheep, and $\geq$ 3.5m for the goitered gazelle. 

\section{Discussion}
\subsection{Discussion on field surveys}
The overall terrain of the study area is flat (as can be seen from Fig.~\ref{Fig6} and Fig.~\ref{Fig8}), except some fragmented rugged areas in the northwest and southeast. In the line transect surveys, the ER values indicated that ibex was only spotted in the first three transects in the east, possibly because this species is known to inhabit mostly the provincial nature reserve, which is southeast to the expressway, and are relatively more active in mountainous terrain~\citep{ibex}. The argali sheep are known to inhabit deserts, grasslands, rocky areas, and shrublands~\citep{Argalisheep}. When compared with the ibex, these sheep might tend to linger more on flat terrain near the location of the expressway. This could be the reason why we found signs of these sheep in a much wider area than those of ibex (indicated by their higher ER values on several transects in the middle and west). In contrast, the high values of goitered gazelle indicate that their demand for crossing activities could be high in the area. Moreover, the abundant signs found during the surveys evidenced that the goitered gazelle has a much larger home range in this area.

The purpose of the belt transect survey was to estimate crossing hotspots along the planned location of the expressway so that the decision on where to put crossing structures can be made. In a general principle, optimal WCSs would very likely be established at the hotspots discovered along the linear infrastructure~\citep{huijser2007wildlife}. However, previous studies have also shown that road kill hotspots might not be the best location to place WCSs~\citep{eberhardt2013road,zimmermann2017road}. In addition, the majority of roads in a desert landscape might act as an attractant for most of the local species, which could then increase animal densities around roads. Also, the increased cover of plants along the expressway embankment and the crossing structures themselves might also alter the pattern of animal movements~\citep{lee2015roads,martinig2017habitat}. Therefore, we speculate that the crossing hotspots and crossing frequencies in the future could be different from what we had estimated. 

During the water surveys, we found ample evidence to support the claim that surface waters are vital in an arid environment~\citep{roshier2008animal,razgour2010pond,zhang2015water,kluever2017influence}. However, the free-standing water on the expressway and nutrient-rich grasses on right-of-ways after heavy rainfall might also be an attractant once the road is in operation~\citep{boarman1997effect,lee2004kangaroo,garland1984effects,starr2001assessing}, which may lead to increasing crossing activities. Again, this will alter the patterns of crossing hotspots in the future. Also, during the tracking surveys, we found that some animal tracks indeed led to the location of the expressway, while some other footprint trajectories led to elsewhere, beyond the study area. This implies that those seven surface waters are visited by animals over a much bigger area.

Most importantly, during the field trips, both our research team and invited experts expressed their concerns about the shrinking habitats and reduced animal populations. Also, no signs of wild Bactrian camel or Mongolian wild ass were found in the area based on our observation. Therefore, we highly recommend that the government should pay more attention to the ecological conditions in northwestern China.

\subsection{Discussion on WCS designs}
The mean distance between estimated crossing hotspots and the nearest WCS is 341m. Considering the vast home range of the ungulate species~\citep{bowman2002dispersal}, as well as a large number of bridges in engineering design (see Appendix A), we believe it should not be difficult for those ungulates to reach an available crossing structure.

Road planning, design, construction, and operation are complex challenges that attempt to balance environmental, economic and social demands~\citep{roberts2015incorporating,huijser2009cost}. Indeed, many conflicts have to be reconciled during the study. We chose underpasses as the passage type for the following reasons: (1) the most abundant species in the area are large- and medium-sized ungulates. Previously reported evidence~\citep{yang2008tibetan,staines2001road} has shown that underpasses are effective for most animals of this size, including red, roe, and fallow deer, while other underpass structures, such as small culverts, can be used by small animals~\citep{clevenger2005performance}; (2) underpasses can be successfully implemented by modifying and adjusting the original bridges in the engineering design to balance costs and benefits. In contrast, overpasses are one type of the most expensive mitigation measures~\citep{huijser2007wildlife}; and (3) it is not practical to build overpasses on the top of bridge structures. Moreover, there are 226 tunnels and small culverts designed in this expressway project. Combining with proposed WCSs, those underpasses can serve multiple species. Some medium-sized and small species, such as fox, hare and reptiles, are known to prefer tunnels and small culverts in the expressway design~\citep{polak2018optimal}. However, small animals might require species-specific WCSs as previous works have found that species' responded to the crossing structures differently~\citep{mcdonald2004elements,clevenger2005performance,ascensao2007factors,martinig2016factors}. Thus, further studies may be needed for those small animals and other species in this project.

Since this WCS design took place in the early stage of the project, the animal movement data was not available for carrying out more rigorous mathematical modeling. Even with a sufficient amount of GPS data, the optimal result of positioning WCSs may still vary due to the bias in high-resolution data~\citep{bastille2018optimizing}. Therefore, a systematic approach, which integrates available data, expert knowledge, and field surveys should be recommended for the decision-making process~\citep{clements2014and,baxter2018turning,boyle2017comparison}.



There are supplementary measures to reduce wildlife-vehicle collisions and provide safe crossing opportunities for ungulates in the area. For example, fences could be recommended to ensure that those ungulates could adapt to the WCSs in the future~\citep{ascensao2007factors,rytwinski2016effective}. Studies have illustrated the importance of the fences~\citep{glista2009review,huijser2016effectiveness}, which could be considered depending on the specific features of the project, such as the degree of road avoidance and traffic volume~\citep{jaeger2004effects}. Moreover, wildlife warning signs are also recommended on the expressway to warn drivers about animals' abnormal crossing activities~\citep{crawford2015drivers,crawford2016drivers}. Moreover, reminder signs are also needed to control the level of traffic noise to prevent animals' avoidance behaviors due to noisy disturbances~\citep{clevenger2000factors}.

The proposed crossing structures will require future monitoring to ensure their effectiveness based on rigorous approach and analysis~\citep{clevenger2003long,ford2009comparison,van2015good}. Because the G7 expressway has been fully operational only since 2017, no monitoring data are yet available. Nevertheless, cameras (with functions in surveillance, night vision, and motion detection) have already been installed in 2017. We will report on the monitoring results in our future work.

Finally, we acknowledge one limitation of this study. The field surveys were conducted mostly during winter time due to the scheduling constraints of the project. This limitation may introduce some bias into the results of surveys~\citep{mata2009seasonal,serronha2013towards}. Therefore, we suggest that future projects should carry out multiple surveys in different seasons in order to cover a comprehensive picture of the periodic features of the target species. 


\section{Conclusion}
We report on a recent road ecology project associated with a regional expressway in Gansu Province to demonstrate how we designed the wildlife crossing structures for the local ungulate species. This paper exemplifies the up-to-date efforts of China in wildlife protections in infrastructure projects.

In summary, the populations of ungulate species in the study area are relatively low. However, we found that the goitered gazelle is relatively more widely and evenly distributed among the local species. We proposed 16 underpass WCSs for this expressway project and the mean deviation distance between them and the nearest hotspots estimated from the field surveys is around 341m. Moreover, the WCSs have a minimum width of 12m and a minimum height of 3.5m. The approach demonstrated in this paper facilitates the practical spatial planning of WCSs and provides insights into designing those crossing structures in a desert landscape. More importantly, we recommend that the government should pay more attention to the ecological conditions in Northwestern China. Finally, future work will focus on analyzing the monitoring data of the proposed crossing structures.

\newpage
\singlespacing
\section*{Fundings}
This research was supported by the Major Scientific Research Project of Shaanxi Academy of Science [grant number: 2016K-04] and the Innovative Research Funds of Tianjin Research Institute for Water Transport Engineering, Ministry of Transport of China [grant number: TKS170206].

\section*{Author contributions}
All of the authors contributed to the conception and design of this study, and have read and approved the final manuscript. 

\section*{Statement of competing interests}
The authors declare no conflicts of interest.

\section*{Acknowledgements}
The authors would like to express their deepest gratitude to all of the volunteers and coordinators who participated in and contributed to the field investigations.

\newpage

\section*{Appendix A}
\begin{table}[hpt!]

\caption{\small Statistical dimensions for bridge designs by engineering team, without regard for wildlife crossings.}
\label{tab1}
\begin{tabular}{|l|c|c|c|c|}
\hline
\centering
Category & \multicolumn{1}{l|}{Number} & \multicolumn{1}{l|}{Mean span (m/column)} & \multicolumn{1}{l|}{Mean height (m)} & \multicolumn{1}{l|}{Mean total length (m)} \\ \hline
large & 7 & 20.00 & 3.64 & 128.86 \\ \hline
Medium & 10 & 20.00 & 3.76 & 70.00 \\ \hline
small & 40 & 10.38 & 2.37 & 17.92 \\ \hline
\end{tabular}
\end{table}

\section*{Appendix B}
\begin{figure}[h!]
\centering
\includegraphics[width=1\textwidth]{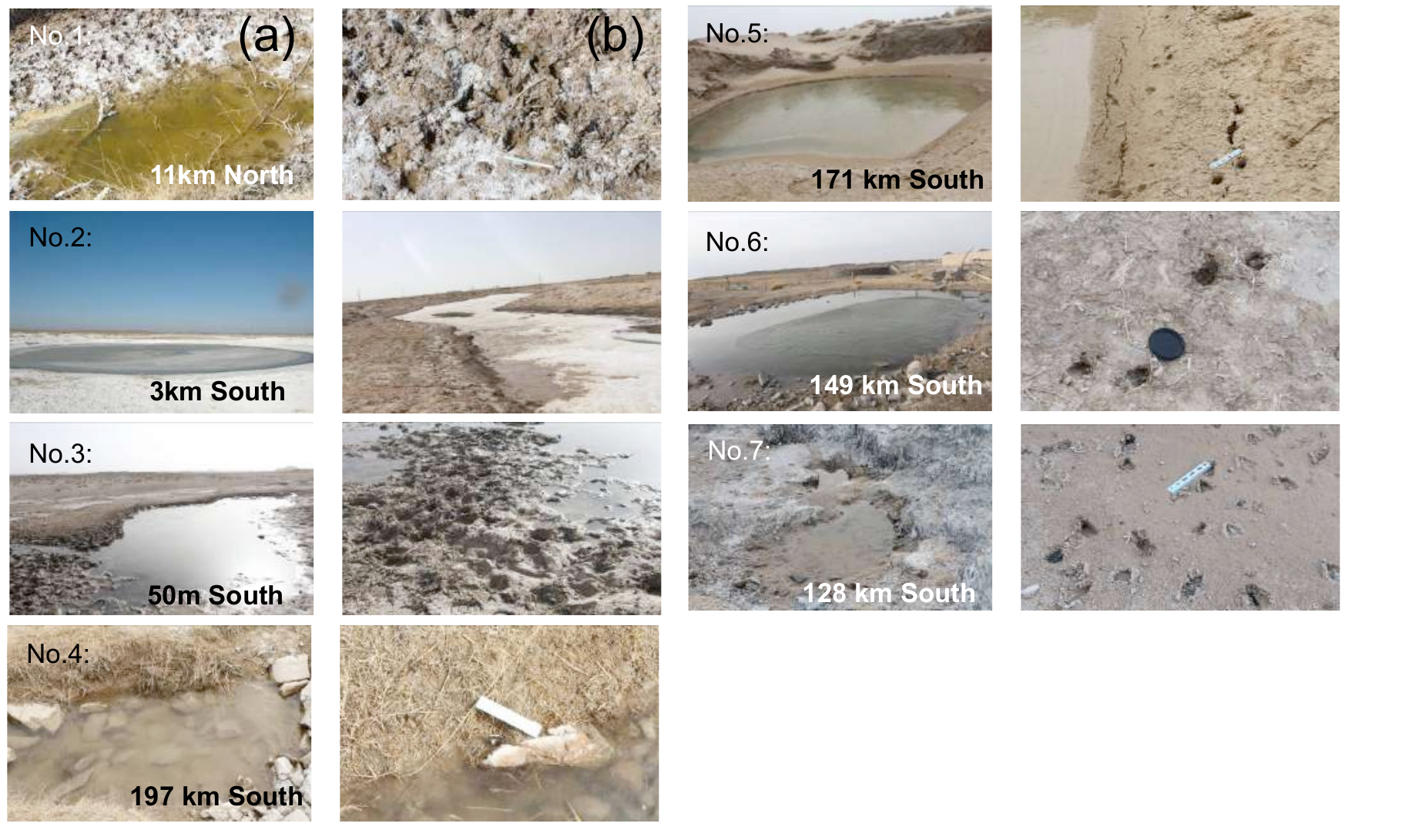}
\caption{\small Surface waters in the study area and their perpendicular locations to the expressway. Photos were taken by surveyors: (a) Field observation of surface waters; (b) Animal signs nearby.  (color is NOT needed)}
\label{Fig7}       
\end{figure}

\newpage


\bibliographystyle{apa}
\bibliography{/Users/junqing/Desktop/1_Papers_data_Codes/Working_papers/Paper_7_Animal_finished/Revision_1/animal_revision}

\end{document}